\documentclass[11pt,a4paper]{article}
\pdfoutput=1
\usepackage{bbm}
\usepackage{amsthm}
\usepackage{amsmath,amssymb}
\usepackage{mathtools}
\usepackage{colortbl}
\usepackage{graphicx}
\usepackage{booktabs}
\definecolor{mygray}{gray}{0.85}
\usepackage{pgfplotstable}
\pgfplotsset{compat= 1.14}
\usepackage{adjustbox}

\DeclareMathOperator{\softmax}{softmax}
\usepackage[]{acl2019}
\usepackage{subfig}
\usepackage{url}
\aclfinalcopy 

\newcommand*{\affaddr}[1]{#1} 
\newcommand*{\affmark}[1][*]{\textsuperscript{#1}}
\newcommand*{\email}[1]{\texttt{#1}}

\newcommand{\sandipan}{\color{blue}}

\title{StRE: Self Attentive Edit Quality Prediction in Wikipedia}
\author{%
Soumya Sarkar\thanks{*Both authors contributed equally}~\affmark[1], Bhanu Prakash Reddy\affmark[*]\affmark[2], Sandipan Sikdar\affmark[3] Animesh Mukherjee\affmark[4] \\ 
\affaddr{IIT Kharagpur, India\affmark[1,2,4],RWTH Aachen, Germany\affmark[3]}\\
\email{soumya015@iitkgp.ac.in\affmark[1],bhanu77prakash@gmail.com\affmark[2]}\\
\email{sandipan.sikdar@cssh.rwth-aachen.de\affmark[3], animesh@cse.iitkgp.ac.in\affmark[4]}
}
\date{}

\begin{document}
\maketitle
\begin{abstract}
  Wikipedia can easily be justified as a behemoth, considering the sheer volume of content that is added or removed every minute to its several projects. This creates an immense scope, in the field of natural language processing toward developing automated tools for content moderation and review. In this paper we propose Self Attentive Revision Encoder (StRE) which leverages orthographic similarity of lexical units  toward predicting the quality of new edits. 
  In contrast to existing propositions which primarily employ features like page reputation, editor activity or rule based heuristics, we utilize the textual content of the edits which, we believe contains superior signatures of their quality. More specifically, we deploy deep encoders to generate representations of the edits from its text content, which we then leverage to infer quality.  
  We further contribute a novel dataset containing $\sim 21M$  revisions across $32K$ Wikipedia pages and demonstrate that \textit{StRE} outperforms existing methods by a significant margin -- at least $17$\% and at most $103$\%. Our pre-trained model achieves such result after retraining on a set as small as $20$\% of the edits in a wikipage. This, to the best of our knowledge, is also the first attempt towards employing deep language models to the enormous domain of automated  content moderation and review in Wikipedia.   

\end{abstract}
\section{Introduction}
Wikipedia is the largest multilingual encyclopedia known to mankind with the current English version consisting of more than $5M$ articles on highly diverse topics which are segregated into categories, constructed by a large editor base of more than $32M$  editors~\cite{hube2019neural}. To encourage transparency and openness, Wikipedia allows anyone to edit its pages albeit with certain guidelines for them\footnote{en.wikipedia.org/Wikipedia:List of policies}. 

\begin{figure}
\centering
    \includegraphics[scale = 0.15]{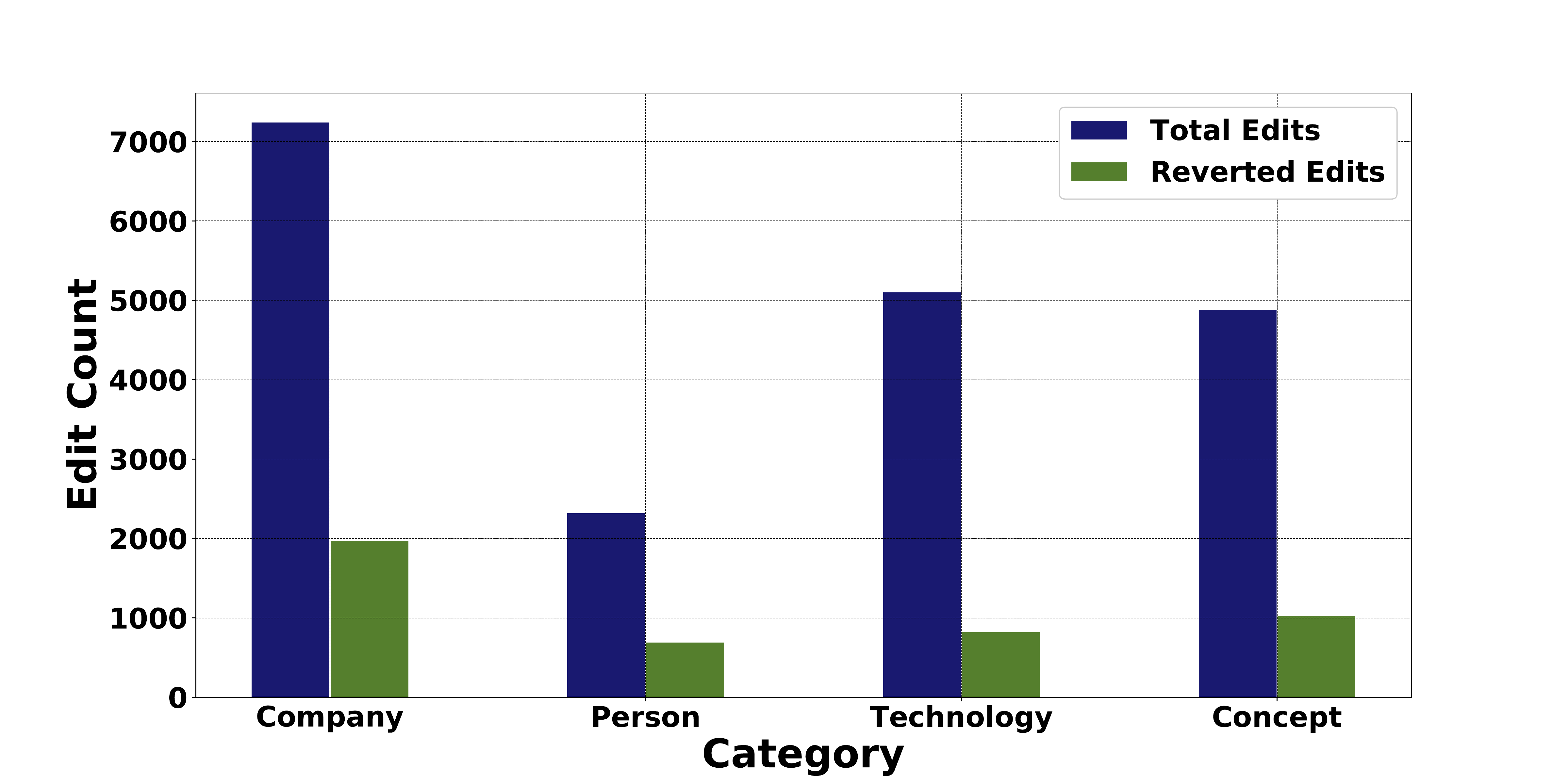}
    \caption{Average number of edits and average number of damaging edits, i.e., reverted edits for four different categories of pages. A fraction (at most $30$\%) of user generated edits are damaging edits.}
    \label{fig:rv_chart}
\end{figure}

\noindent\textbf{Problem}: The inherent openness of Wikipedia has also made it vulnerable to external agents who intentionally attempt to divert the unbiased, objective discourse to a narrative which is aligned with the interest of the malicious actors. 
Our pilot study on manually annotated $100$ Wikipedia pages of four categories ($25$ pages each category) shows us that at most $30$\% of the edits are reverted (See Fig~\ref{fig:rv_chart}). Global average of number of reverted damaging edits is $\sim 9$\%\footnote{stats.wikimedia.org/EN/PlotsPngEditHistoryTop.htm}. This makes manual intervention to detect these edits with potential inconsistent content, infeasible. 
 Wikipedia hence deploys machine learning based classifiers \cite{west2010stiki,halfaker2015artificial} which  primarily leverage hand-crafted features from three aspects of revision (i) basic text features like repeated characters, long words, capitalized words etc. (ii) temporal features like inter arrival time between events of interest (iii) dictionary based features like presence of any curse words or informal words (e.g.,  `hello', `yolo'). Other feature based approaches include ~\cite{daxenberger2013automatically,bronner2012user} which generally follow a similar archetype. 

\begin{figure}
\centering
    \includegraphics[width=3.0in]{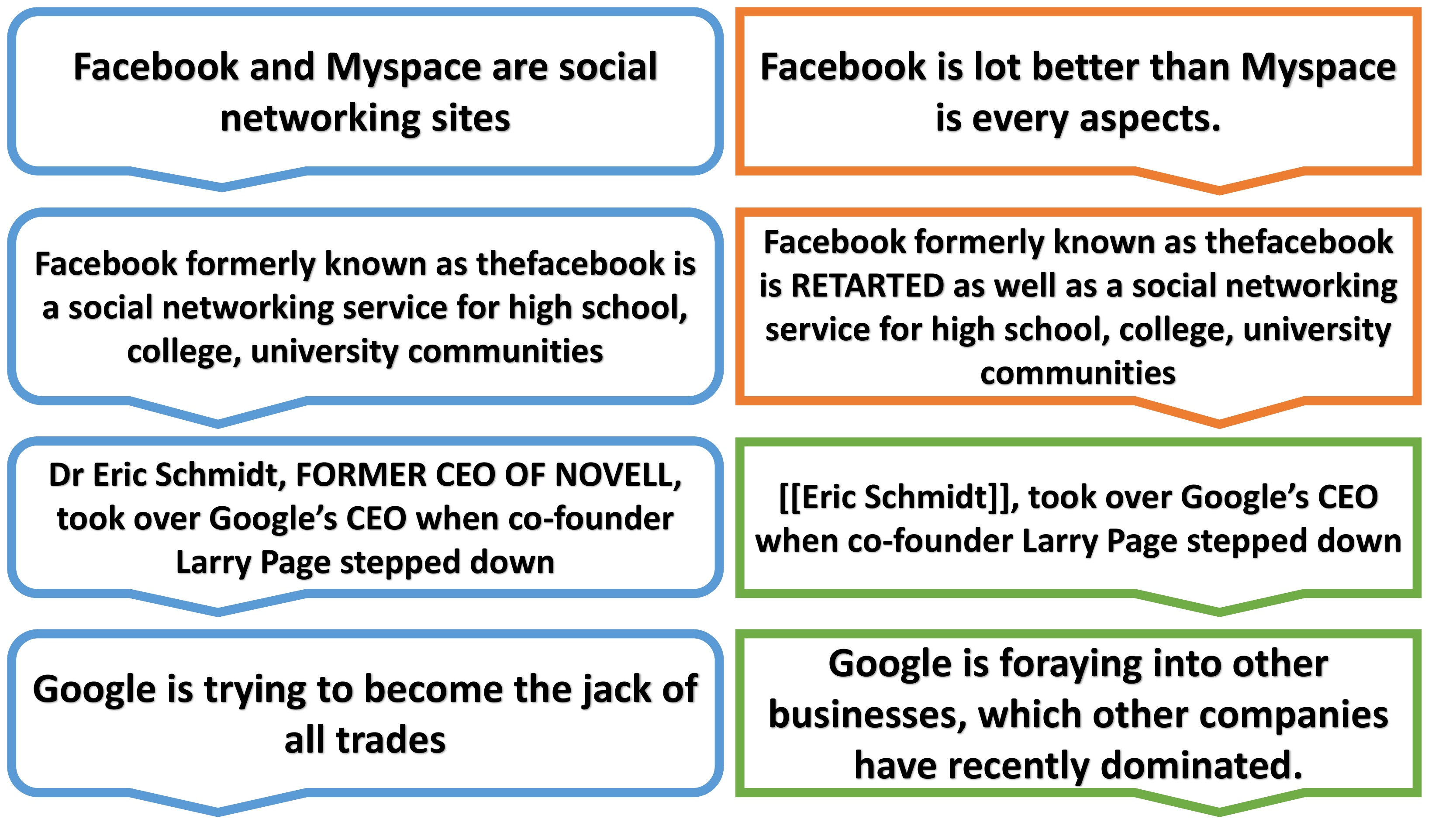}
    \caption{Examples of edits in {\it Facebook} and {\it Google} Wikipedia page. The \textcolor{blue}{blue} bubbles are the original sentences. The \textcolor{orange}{orange} bubbles indicate damaging edits while the \textcolor{green}{green} bubbles indicate `good faith' edits. Good faith edits are unbiased formal English sentence while damaging edits often correspond to incoherent use of language, abusive language, imperative mood, {\sandipan opinionated} sentences etc. }
    \label{fig:edit_example}
\end{figure}

\noindent\textbf{Proposed model}: In most of the cases, the edits are reverted because they fail to abide by the edit guidelines, like usage of inflammatory wording, expressing opinion instead of fact among others (see Fig~\ref{fig:edit_example}).~These flaws are fundamentally related to the textual content rather than temporal patterns or editor behavior that have been deployed in existing methods. Although dictionary based approaches do look into text to a small extent (swear words, long words etc.), they account for only a small subset of the edit patterns. We further hypothesize  that owing to the volume and variety of Wikipedia data, it is impossible to develop a feature driven approach which can  encompass the wide array of dependencies present in text. 
In fact, we show that such approaches are inefficacious in identifying most of the damaging edits owing to these obvious limitations.
We hence propose {\bf S}elf A{\bf t}ttentive {\bf R}evision {\bf E}ncoder (StRE) which extracts rich feature representations of an edit that can be further utilized to predict whether the edit has damaging intent.
 In specific, we use two stacked recurrent neural networks to encode the semantic information from sequence of characters and sequence of words which serve a twofold advantage. While character embeddings extract information from out of vocabulary tokens, i.e., repeated characters, misspelled words, malicious capitalized characters, unnecessary punctuation etc., word embeddings extract meaningful features from curse words, informal words, imperative tone, facts without references etc. We further employ attention mechanisms~\cite{bahdanau2014neural} to quantify the importance of a particular character/word. Finally we leverage this learned representation to classify an edit to be damaging or valid. Note that {\em StRE} is reminiscent of structured self attentive model proposed in ~\cite{lin2017structured} albeit used in a different setting.

\noindent\textbf{Findings}: To determine the effectiveness of our model, we develop an enormous dataset consisting of $\sim 21M$ edits across $32K$ wikipages.
We observe that {\em StRE} outperforms the closest baseline by at least 17\% and at most 103\% in terms of AUPRC. 
Since it is impossible to develop an universal model which performs equally well for all categories, we develop a transfer learning~\cite{howard2018universal} set up which allows us to deploy our model to newer categories without training from scratch. This further allows us to employ our model to pages with lower number of edits. 

\noindent{\bf Contributions}: Our primary contributions in this paper are summarized below - \\ 
    (i) We propose a deep neural network based model to predict edit quality in Wikipedia which utilizes language modeling techniques, to encode semantic information in natural language. \\
    (ii) We develop a novel dataset consisting of $\sim 21M$ unique edits extracted from $\sim 32K$ Wikipedia pages. In fact our proposed model outperforms all the existing methods in detecting damaging edits on this dataset. \\
    (iii) We further develop a transfer learning set up which allows us to deploy our model to newer categories without the need for training from scratch. 
    
    Code and sample data related to the paper are available at \url{https://github.com/bhanu77prakash/StRE}.

\section{The Model}
\begin{figure}
\centering
    \includegraphics[width=3.0in]{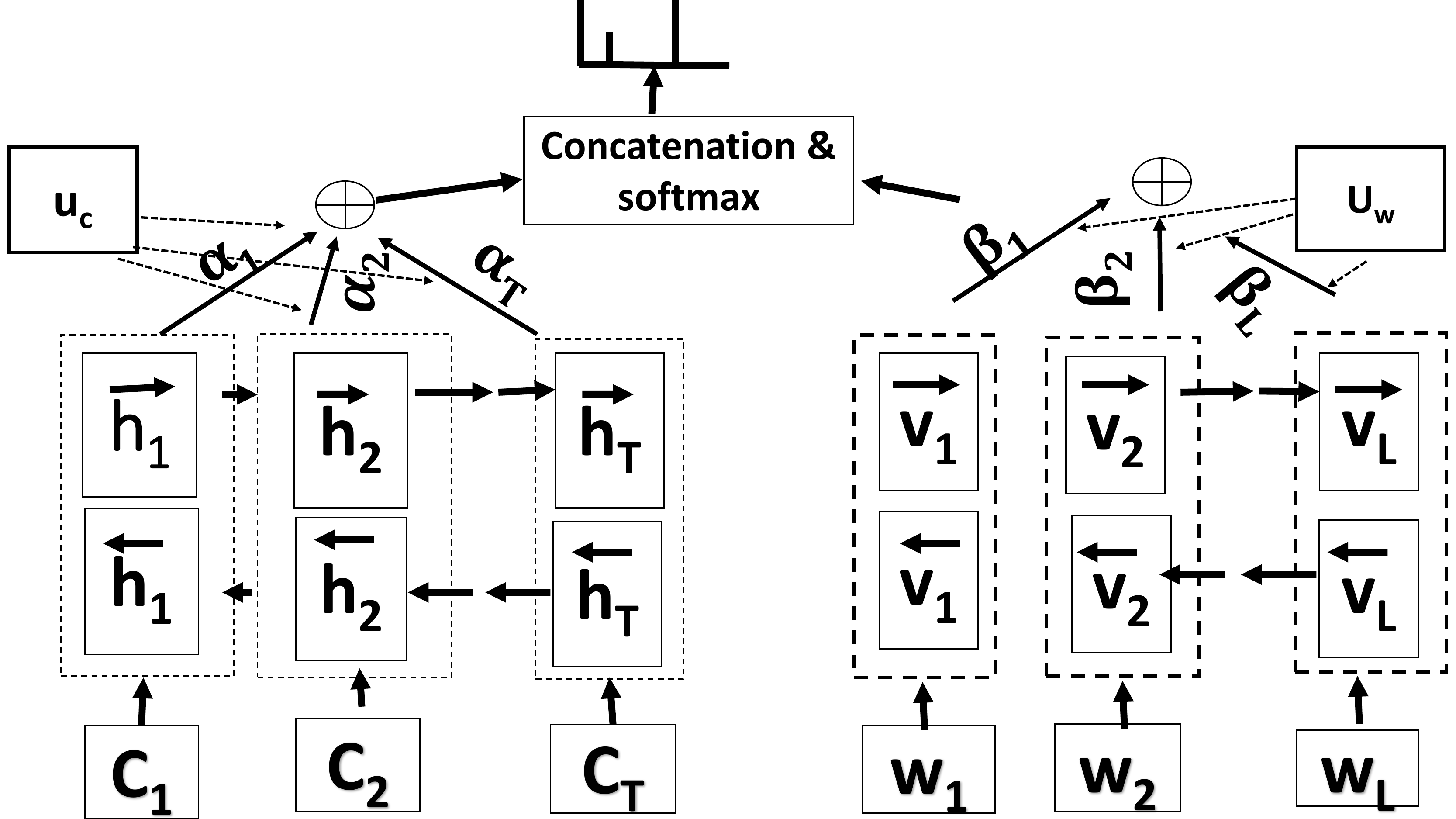}
    \caption{Overall architecture of StRE. The character encoding and the word encoding components of the model are shown in the left and right respectively. This is followed by the attention layer followed by concatenation and \textit{softmax}.}
    \label{fig:sample_model}
\end{figure}

In this section we give a detailed description of our model. 
We consider an edit to be a pair of sentences with one representing the original ($P_{or}$) while the other representing the edited version ($P_{ed}$). The input to the model is the concatenation of $P_{or}$ and $P_{ed}$ (say $P = \{P_{or}||P_{ed}\}$) separated by a delimiter (`$||$'). We assume $P$ consists of $w_i$ words and $c_i$ characters. Essentially we consider two levels of encoding - (i) \textbf{character level} to extract patterns like repeated characters, misspelled words, unnecessary punctuation etc. and (ii) \textbf{word level} to identify curse words, imperative tone, opinionated phrases etc. In the following we present how we generate a representation of the edit and utilize it to detect malicious edits. The overall architecture of {\em StRE} is presented in Fig~\ref{fig:sample_model}.

\subsection{Word encoder}
Given an edit $P$ with $w_i, i\in [0,L] $ words, we first embed the words through a pre-trained embedding matrix $W_e$ such that $x_i = W_e w_i$.  
This sequence of embedded words is then provided as an input to a bidirectional LSTM~\cite{hochreiter1997long} which provides representations of the words by summarizing information from both directions. 
\begin{align}
      x_i &= W_e w_i,\, i \in [0,L] \\
   \overrightarrow{v}_i &= \overrightarrow{LSTM}(x_i),\, i \in [0,L] \\
   \overleftarrow{v}_i &= \overleftarrow{LSTM}(x_i),\, i \in [L,0] 
\end{align}

We obtain the representation for each word by concatenating the forward and the backward hidden states $v_i = [\overrightarrow{v}_i,\overleftarrow{v}_i]$. Since not all words contribute equally to the context, we deploy an attention mechanism to quantify the importance of each word. The final representation is then a weighted aggregation of the words. 
\begin{align}
 u_i &= \sigma(W_w v_i + b_w) \\
 \beta_i &= \frac{exp(u_i^T u_w)}{\sum_{i=0}^{T}exp(u_i^T u_w)} \\
 R_w &= \sum_i \beta_i v_i
\end{align}

To calculate attention weights ($\alpha_i$) for a hidden state $h_i$, it is first fed through a single layer perceptron and then a \textit{softmax} function is used to calculate the weights. Note that we use a word context vector $u_w$ which is randomly initialized and is learnt during the training process. 
The use of context vector as a higher level representation of a fixed query has been argued in \cite{sukhbaatar2015end,kumar2016ask}. Note that the attention score calculation is reminiscent of the one proposed in \cite{yang2016hierarchical}.

\subsection{Character encoder}
The character encoder module is similar to the word encoder module with minor differences. Formally we consider $P$ ($\{P_{or}||P_{ed}\}$) as a sequence of $T$ characters $c_i$, $i \in [0,T]$. Instead of using pre-trained embeddings as in case of word encoder, we define an embedding module, parameters of which is also learned during training which is basically an MLP. Each embedded character is then passed through a bidirectional LSTM to obtain the hidden states for each character. Formally, we have 
\begin{align}
    y_i &= \sigma(W_c c_i + b_c),\, i \in [0,T] \\
   \overrightarrow{h}_i &= \overrightarrow{LSTM}(y_i),\, i \in [0,T] \\
   \overleftarrow{h}_i &= \overleftarrow{LSTM}(y_i),\, i \in [T,0]
\end{align}

We next calculate the attention scores for each hidden state $h_i$ as 
\begin{align}
 z_i &= \sigma(W_c h_i + b_c) \\
 \alpha_i &= \frac{exp(z_i^T u_c)}{\sum_{i=0}^{T}exp(z_i^T u_c)} \\
 R_c &= \sum_i \alpha_i h_i
\end{align}

Note that $u_c$ is a character context vector which is learned during training.


\subsection{Edit classification}

The edit vector $E_p$ (for an edit $P$) is the concatenation of character and word level encodings $E_p = [R_c,R_w]$ which we then use to classify whether an edit is valid or damaging. Typically, we perform 
\[p = softmax(W_p E_p + b_p)\]
Finally we use binary cross entropy between predicted and the true labels as our training loss.

\subsection{Transfer learning setup}

Note that it is not feasible to train the model from scratch every time a new page in an existing or a new category is introduced. Hence we propose a transfer learning setup whereby, for a new page, we use the pre-trained model and only update the weights of the dense layers during training. The advantages are twofold - (i) the model needs only a limited amount of training data and hence can easily be trained on the new pages and (ii) we benefit significantly on training time.

\section{Dataset}

\begin{figure*}[t]
    \begin{minipage}[b]{\textwidth}
  	\centering
    \includegraphics[scale=0.15]{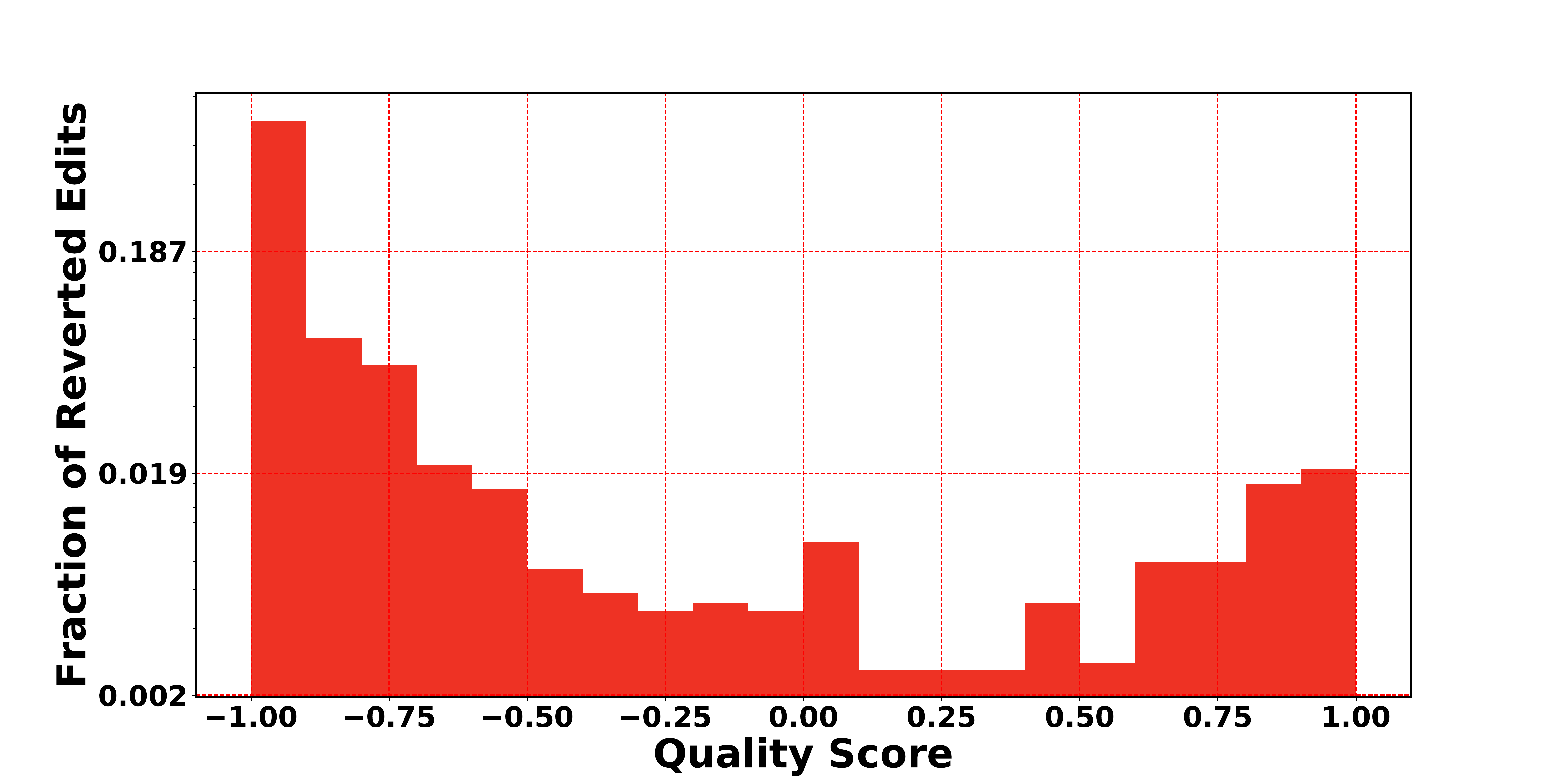}
     \includegraphics[scale=0.15]{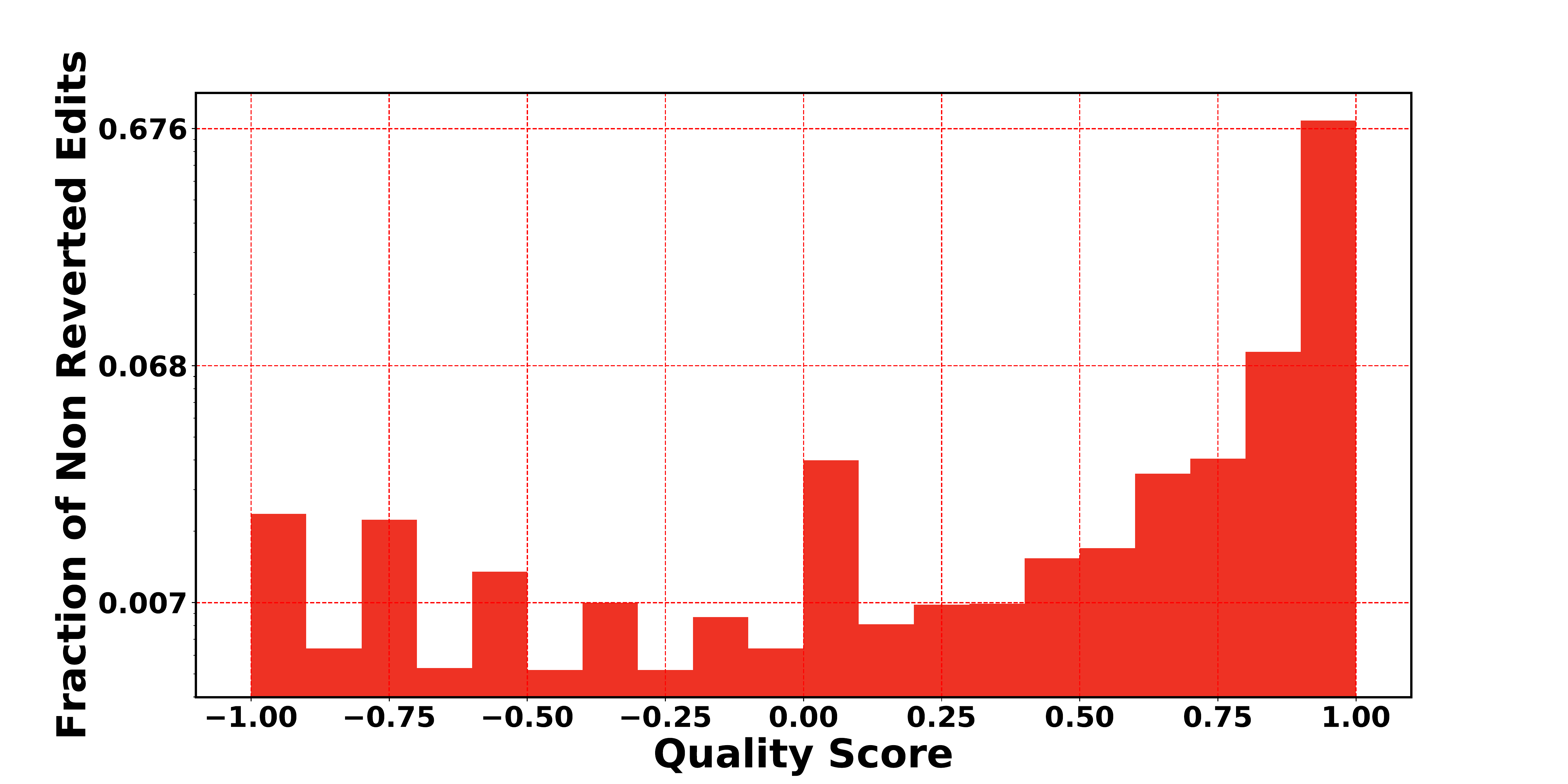}
   \end{minipage}
   \caption{Distribution of the quality score for the revert edits (left) and non-reverted edits(right). The $y$-axis is in log scale. The plot shows that a large proportion of low quality edits are reverted and a large proportion of high quality edits are not reverted; hence this observation acts as a validation for our quality score metric.}
   \label{fig:edit_quality}
\end{figure*}

Wikipedia provides access to all Wikimedia project pages in the form of xml dumps, which is periodically updated\footnote{https://dumps.wikimedia.org/enwiki/20181120}. We collect data from dumps made available by English Wikipedia project on June 2017 which contains information about $5.5M$ pages. 


We extract a subset of pages related to the Computer Science category in Wikipedia. 
Utilizing the category hierarchy\footnote{Dbpedia.org}~\cite{auer2007dbpedia} (typically a directed graph containing parent and child categories), 
we extract all articles under the Computer Science category up to a depth of four levels which accounts for $48.5K$ Wikipedia pages across $1.5K$ categories\footnote{Typical categories include `Computational Science', `Artificial Intelligence' etc.}.
We filter out pages with at least $100$ edits which leaves us with $32K$ pages. 
For each page in our dataset we performed pairwise difference operation\footnote{https://docs.python.org/2/library/difflib.html} between its current and previous versions to obtain a set of pairs with each consisting of a sentence and its subsequent modified version.  

\noindent{\bf Edit quality}: 
In order to train our model to identify quality edits from damaging edits we need a deterministic score for an edit. 
Our quality score is based on the intuition that if changes introduced by the edit are preserved, it signals that the edit was beneficial, whereas if the changes are reverted, the edit likely had a negative effect. This idea is adapted from previous work of~\citet{adler2011wikipedia}.

Consider a particular article and denote it by $v_k$ its $k$-th revision (i.e., the state of the article after the $k$-th edit).
Let $d(u, v)$ be the Levenshtein distance between two revisions. 
We define the \emph{quality} of edit $k$ from the perspective of the article's state after $\ell \ge 1$ subsequent edits as
$$q_{k \mid \ell} = \frac{d(v_{k-1}, v_{k + \ell}) - d(v_k, v_{k + \ell})}{d(v_{k-1}, v_k)}.$$

Intuitively, the quantity $q_{k \mid \ell}$ captures the proportion of work done on edit $k$ that remains in revision $k + \ell$ and it varies between  $q_{k \mid \ell} \in [-1, 1]$, when the value
falls outside this range, it is capped within these two values. 
We compute the mean quality of the edit by averaging over multiple future revisions as follows

$$q_k = \frac{1}{L} \sum_{\ell = 1}^L q_{k \mid \ell}$$

where $L$ is the minimum among the number of subsequent revisions of the article. We have taken $L=10$, which is consistent with the previous work of~\citet{yardim2018can}.

\noindent{\bf Edit label}:
For each pair of edits we compute the edit quality score. If quality score is $\ge 0$ we label an edit to be $-1$, i.e., done in good faith. However all edits with quality score $< 0$ are labeled 1, i.e., damaging edits. We further check that bad quality edits are indeed damaging edits by calculating what fraction of low score edits are reverted and what fraction of high score edits are not reverted. This result is illustrated in Figure~\ref{fig:edit_quality}. Information whether an edit is reverted or not can be calculated by mining Wikipedia's revert graph following the same technique illustrated by~\cite{kittur2007he}. The results clearly show that a large proportion of bad quality edits are indeed reverted by the editors and similarly a large fraction of good quality edits are not reverted. Though bad quality edits are often reverted, all reverted edits are not bad. Malicious agents often engage in interleaving reverts, i.e., edit wars~\cite{kiesel2017spatio} as well as pesudo reverts. Hence we use quality metric to label damaging edits which is well accepted in the literature~\cite{adler2008measuring}. We provide a summary of the data in Table~\ref{tab:wikidata}. Our final data can be represented by a triplet $<s_i,s_f,l>$ where $s_i$ is the initial sentence, $s_f$ is the modified sentence and $l$ indicates the edit label.

\begin{table}
  \centering
  \scalebox{0.75}{
  \begin{tabular}{lc}
    \toprule
    Resources & Count \\
    \midrule
    Pages  & 32394 \\
    Total edits & 21,848960\\
    Positive edits &  15,791575 \\
    Negative edits &  6,057385\\
    \bottomrule
  \end{tabular}}
  \caption{Summary of the dataset.}
  \label{tab:wikidata}
\end{table}
\section{Experiments}\label{Sec:exp_setup}





In this section we demonstrate the effectiveness of our model compared to other existing techniques. Typically, we consider two sets of experiments - 
(i) category level and (ii) page level. 
In category level experiments (see section~\ref{sec:exp_cat}) we first form a random sample of  data points belonging to pages in a fixed category. Our objective is to first train on edits related to a fixed page category and test on new edits belonging to pages of the same category. We further show through rigorous experiments that existing approaches of transfer learning and fine tuning~\cite{howard2018universal} can be applied to increase the efficacy of our approach.  In page level experiments in section~\ref{sec:exp_page}, we abandon the category constraint (as in case of category level experiments) and train (test) on edits irrespective of the category of the page which it belongs to and demonstrate that our model is equally effective. 


\subsection{Baseline approaches}
We use two variants of our proposed model -- word embedding with attention ({\it Word+Att}), character embedding with attention ({\it Char+Att}) as two baselines to compare to our model. We also compare existing feature based and event based approaches for edit quality prediction. We give a brief description of the other state-of-the-art baselines in the subsequent subsections.
\subsubsection{ORES}
The Objective Revision Evaluation Service (ORES)~\cite{Wiki} is a web service developed by Wikimedia foundation that provides a machine learning-based
scoring system for edits. More specifically, given an edit,
ORES infers whether an edit causes damage using linguistic features and edit based features (e.g., size of the revision etc.)


\subsubsection{ORES++}
In order to make it more competitive, we further augment ORES by adding linguistic quality indicators as additional features obtained from the {\it Empath} tool~\citet{fast2016empath}. This tool scores edits on $16$ lexical dimensions such as `ugliness', `irritability', `violence' etc. We also use the count of POS tags following~\citet{manning2014stanford} as well as the count of mispelled words as features using aspell dictionary~\citet{atkinson2006gnu}.

\subsubsection{Interrank}
Interrank~\cite{yardim2018can} is a recent quality-prediction method which does not use any explicit content-based features but rather predicts quality of an edit by learning editor competence and page reputation from prior edit actions. The performance of Interrank has been revealed to be very close to ORES.

\subsection{Model configuration}
We use 300 dimensional pre-trained word Glove vector~\cite{pennington2014glove} and 300 dimensional ASCII character embedding~\cite{minimaxir}. We also use 64 dimensional hidden layer in our model, followed by attention layer and three stacks of dense layer. Our context vector in the attention layer is of 64 dimensions and dense layers are 256, 64 and 16 dimensions. We further utilize dropout probability of 0.5 in the dense layers. We also employ binary cross entropy as loss function and Adam~\cite{kingma2014adam} optimizer with learning rate $0.01$ and  weight decay of  $0.0001$ to train our model. The batch size is set to 250. 


\subsection{Category level experiments}\label{sec:exp_cat}
In this set of experiments we essentially train and test on pages in the same category.  
\subsubsection{Page specific model}
As a first step towards determining the potential of our model, we train our model on a set of edits of a particular page and predict on the rest. To this aim we manually annotate top 100 pages in terms of total number of edits, into four categories, i.e., company, concept, technology, and person. Such granular level category annotation is not available from Wikipedia hierarchy which directs us towards annotation. In each category we tabulate the count of positive and negative datapoints in Table~\ref{tab:count}). For each page we randomly select $80$\% edits for training, $10$\% edits for validation and $10$\% edits as held out set. We train our model on $80$\% and tune it on the validation set. We finally test on the held out set. The same procedure is followed for {\em Word+Att}, {\em Char+Att}, {\em ORES++}. For all these models the AUPRC (mean,std) across all pages are presented in Table~\ref{tab:result1}. Since {\em ORES} is already a pre-trained model we test on the combined held out set of the pages. Note that {\em Interrank} is not designed for page level training and further requires large training data.
Hence, for {\em Interrank} we train on the combined training set of the pages and test on the combined held out set.  Results obtained on the held out set are reported in Table~\ref{tab:result1}. Our experiments clearly show that {\em StRE} outperforms baselines by a significant margin (at least $10$\%). We also see that individual components of our model, i.e., {\em Char+Att} and {\em Word+Att} do not perform as well as {\em StRE} which further validates our architecture. Moreover, {\em Interrank} performs poorly despite combined dataset which shows that language modelling is essential in edit quality prediction. 
\begin{table}[bht]
\centering
{\scriptsize
\begin{tabular}{|c|c|c|c|c|}
\hline
 Edits & \multicolumn{1}{c|}{\textbf{Company}} & \multicolumn{1}{c|}{\textbf{Concept}} & \multicolumn{1}{c|}{\textbf{Technology}} & \multicolumn{1}{c|}{\textbf{Person}} \\ \hline
\textbf{+ve examples} 
 &   813400 & 227308 & 294125 &  79035 \\ \hline 
{\bf -ve Examples} & 649078 & 124323 & 169091 & 28505 \\ \hline
{\bf Total examples} & 1462478 & 351631 & 463216 & 107540  \\ \hline
\end{tabular}
}
\caption{Total number of data points along with positive and negative samples for the top five pages in terms of edit count in each category.}
\label{tab:count}
\end{table}

\begin{table*}[bht]
\centering
{\scriptsize
\begin{tabular}{|c|c|c|c|c|}
\hline
 & \multicolumn{1}{c|}{\textbf{Company}} & \multicolumn{1}{c|}{\textbf{Concept}} & \multicolumn{1}{c|}{\textbf{Technology}} & \multicolumn{1}{c|}{\textbf{Person}} \\ \hline
\textbf{Models} 
 &  {\bf AUPRC} & {\bf AUPRC} & {\bf AUPRC} &  {\bf AUPRC} \\ \hline 
{\em ORES} & 0.72 & 0.76 & 0.71 & 0.63 \\ \hline
{\em ORES++} &  $0.84 \pm 0.03 $ & $ 0.85 \pm 0.03$ & $0.87 \pm 0.02$ & $0.85 \pm 0.03$ \\ \hline
{\em Interrank} & 0.35 &  0.47 & 0.42 & 0.38  \\ \hline
{\em Word+Att} &  $0.63 \pm 0.02 $ & $0.74 \pm 0.03$ &  $0.72 \pm 0.01$ & $0.78 \pm 0.02 $ \\ \hline
{\em Char+Att} &  $0.91 \pm 0.01$ & $0.84 \pm 0.02 $ & $0.83 \pm 0.02$ & $0.81 \pm 0.02$
 \\ \hline
{\em StRE} & \cellcolor{mygray}{$0.95 \pm 0.02 $} & \cellcolor{mygray}{$0.89\pm 0.01$} & \cellcolor{mygray}{$0.91 \pm 0.01$} & \cellcolor{mygray}{$0.87 \pm 0.02$}
\\ \hline
\end{tabular}
}
\caption{\textsc{AUPRC} scores, with the best results in bold and gray background on the annotated dataset.}
\label{tab:result1}
\end{table*}

\subsubsection{New page: same category}
We now explore a more challenging setup for our model whereby instead of training and testing on edits of a specific annotated category, we train on edits of pages of a particular category but test on a previously unseen (during training) page of the same category. Specifically, for a given category, we train our model on 90\% of pages and test our models on unseen page edits in the same category from our annotated dataset. The obtained results are tabulated in Table~\ref{tab:transfer}(a). Results show that such an approach is indeed fruitful and can be applied on pages which has very few edits utilizing intra-category pages with large edit counts. 

\noindent\textbf{Transfer learning results}: Our results can be further improved by applying ideas of transfer learning. For each  new page, we can initialize our model by pre-trained weights learned from training on other intra-category pages. We can then train the dense layer with only 20\% of new datapoints randomly selected from the new page and test on the remaining $80$\%. This approach is adapted from the state-of-the-art transfer learning approaches~\cite{howard2018universal,dehghani2017learning} where it has been shown to work on diverse NLP tasks. Such an approach achieves at least $3$\% and at most $27$\% improvement over prior results.

\begin{table*}[bht]
\subfloat[][Intra category \textsc{AUPRC}.]{
\
\scalebox{0.55}{
\begin{tabular}{|c|c|c|}
\hline
{\bf Category} & \begin{tabular}{@{}c@{}} {\bf Testing without} \\ {\bf Retraining}\end{tabular}& 
\begin{tabular}{@{}c@{}} {\bf Testing with} \\ {\bf $20$\% Retraining}\end{tabular}  \\
\hline
Person & 0.81 & \cellcolor{mygray}{0.85} \\ \hline
Concept & 0.77 & \cellcolor{mygray}0.91 \\ \hline
Company & 0.76 & \cellcolor{mygray}0.88 \\ \hline
Technology & 0.68 & \cellcolor{mygray}0.88 \\
\hline
\end{tabular}}}
\qquad
\subfloat[][Inter category \textsc{AUPRC}.]{
\scalebox{0.55}{
\begin{tabular}{|c|c|c|}
\hline
{\bf Category} & \begin{tabular}{@{}c@{}} {\bf Testing without} \\ {\bf Retraining}\end{tabular}& 
\begin{tabular}{@{}c@{}} {\bf Testing with} \\ {\bf $20$\% Retraining}\end{tabular}  \\
\hline

Person &0.67 & \cellcolor{mygray}0.82\\
\hline
Concept & 0.63 & \cellcolor{mygray}0.81 \\ \hline
Company & 0.71 & \cellcolor{mygray}0.82 \\ \hline
Technology & 0.72 & \cellcolor{mygray}0.89 \\ 
\hline
\end{tabular}}}
\qquad
\subfloat[][Category agnostic \textsc{AUPRC}.]{
\scalebox{0.55}{
\begin{tabular}{|c|c|c|}
\hline
{\bf Category} & \begin{tabular}{@{}c@{}} {\bf Testing without} \\ {\bf Retraining}\end{tabular}& 
\begin{tabular}{@{}c@{}} {\bf Testing with} \\ {\bf $20$\% Retraining}\end{tabular}  \\
\hline
Person & 0.71 & \cellcolor{mygray}0.83\\
\hline
Concept & 0.85 & \cellcolor{mygray}0.90 \\
\hline
Company & 0.74 & \cellcolor{mygray}0.86 \\
\hline
Technology & 0.77 & \cellcolor{mygray}0.84 \\ \hline
\end{tabular}}}
\caption{Results for intra-category, inter-category and category agnostic predictions without and with transfer learning. The transfer learning approach is always beneficial.}
\label{tab:transfer}
\end{table*}

\subsubsection{New page: different category}
We now probe into how our model performs when tested on a page belonging to a previously unseen category. As a proof of concept, we train on all pages belonging to three categories (inter-category training) and test on a new page from the fourth category. We perform this experiment considering each of the four categories as unknown one by one in turn. The obtained results are presented in Table~\ref{tab:transfer}(b). Clearly, the results are inferior compared to intra-category training which corroborates with our argument that different category of pages have unique patterns of edits. 

\noindent\textbf{Transfer learning results}: However, we alleviate the above problem by utilizing transfer learning approaches. In specific, we initialize our model with weights pre-trained on inter-category pages and train only the final dense layer on $20$\% of the new edits from the fourth category. Results point that we can obtain significant improvements, i.e., at least $10$\% and at most $28$\%. This is very a promising direction to pursue further investigations, since it is very likely that abundant edits may be present in distant categories while very limited edits may manifest in a niche category that has low visibility.

\subsubsection{Multi category training}\label{sec:exp_page}
Finally, we proceed toward a category agnostic training paradigm. Essentially, we hold out 10\% pages of the annotated set for each category. We train on all remaining pages irrespective of the category information and test on the held out pages from each category. We report the  results in Table~\ref{tab:transfer}(c). Since our model learns from edits in all category of pages, we are able to obtain better results from inter category setup.  We further employ transfer learning (as in previous sections) on the new page which improves the results significantly (at least 6\% and at most $16\%$). 

To summarize the results in this section, we observe that testing on a previously unseen category leads to under-performance. However, retraining the dense layers with a few training examples drastically improves the performance of our model.  

\subsection{Page level experiments}\label{sec:exp_page}
We now consider an experimental setup agnostic of any category. 
In this setting, to train our model we form a set of edits which comprises $20$\% of our total edits in the dataset. This edits are taken from the pages which have largest edit count. Quantitatively, we impose a cap on the total number of edits to be $20$\% of the entire edit count. Subsequently, we start pooling training data from the largest page, followed by the second largest page and so on until our budget is fulfilled. The whole data so accumulated is divided into 80\% training, 10\% validation and 10\% test sets.
Results on this $10$\% held out data are reported in Table~\ref{tab:final} as training AUPRC. We compare our model against other text based and event based quality predictor baselines. Since {\em ORES} is an already pre-trained web based service, we obtained \textsc{AUPRC} on the $10$\% held out set. In case of {\it Interrrank}, $90$\% of the data is used for training and $10$\% is used as held out set (as reported in the paper~\cite{yardim2018can}). Results show that our model performs significantly better than the baselines (by $24$\% in case of {\it ORES} and by $131$\% in case of {\em Interrank}). 

\noindent\textbf{Transfer learning results}: For each of the remaining pages in our data we first utilize our pre-trained model from the last step. However,  
we train the dense layers with randomly selected $20$\% datapoints from the page to be tested. The remaining data is used for testing. We follow this procedure for all remaining pages and calculate the mean test \textsc{AUPRC} along with standard deviation which we report in Table~\ref{tab:final}. In case of {\em ORES} we evaluate on the $80$\% data. In case of {\em Interrrank}, we merge all remaining data into a single dataset and use $90$\% of the data for training and $10$\% for test. We show that transfer learning approach can be useful in this setting and we obtain $17$\% improvement compared to {\em ORES} and $103$\% improvement compared to {\em Interrank}. 

\begin{table}[]
    \centering
    \scalebox{0.75}{
\begin{tabular}{|c|c|c|}
\hline
{\bf Model} & \begin{tabular}{@{}c@{}} {\bf Training} \\ {\bf AUPRC}\end{tabular}& 
\begin{tabular}{@{}c@{}} {\bf Testing} \\ {\bf AUPRC}\end{tabular}  \\
\hline
ORES & $0.77$ & $0.75$\\
\hline
Interrank & $0.41$ & $0.42$ \\
\hline
{\em Word+Att} & $0.64 $ & $ 0.77 \pm 0.1$ \\
\hline
{\em Char+Att} & $0.92$ & $0.83 \pm 0.09$\\
\hline
StRE & \cellcolor{mygray}0.95 & \cellcolor{mygray}$0.88 \pm 0.09$ \\
\hline
\end{tabular}}
    \caption{Comaprison between {\it StRE} and baselines on complete dataset.}
    \label{tab:final}
\end{table}






\section{Discussion}
\begin{table*}
  \centering
  \scalebox{0.75}{
  \begin{tabular}{c  c}
    \toprule
    Original version & Revised version\\
    \midrule
    \begin{tabular}{@{}l@{}}
         Google Maps offers detailed {\it \textcolor{blue}{streetmaps}} \\ and {\it \textcolor{blue}{route planning}} information. \\
    \end{tabular} &
    \begin{tabular}{@{}l@{}}
        Google Maps offers detailed {\it \textcolor{blue}{streetmaps}} \\ 
         and {\it \textcolor{blue}{route planning}} information  in \\ 
         United States and Canada. \\
    \end{tabular}  
    \\
    \midrule
    \midrule
    \begin{tabular}{@{}l@{}}
         Proponents argued that \it {\textcolor{blue}{privacy complaints}} \\ are baseless. \\
    \end{tabular} &
    \begin{tabular}{@{}l@{}}
         Proponents of  {\it \textcolor{blue}{trusted computing}} argue that \\ {\it \textcolor{blue}{privacy complaints}}
         have  been  addressed in \\ the existing  {\it \textcolor{blue}{specifications}} - possibly as a \\ result of   criticism  of early  versions of  the \\ {\it \textcolor{blue}{specifications}}. 
    \end{tabular}
    \\
    \bottomrule
    \end{tabular}}
  \caption{Anecdotal examples of edits in {\it Google Maps} and {\it Internet Privacy} wikipage. Here the general model fails to identify negative examples while retraining the dense layer learns better representations and identifies the negative examples correctly. Page specific tokens are colored in \textcolor{blue}{blue}.}
    \label{tab:transfer_example}
\end{table*}

\begin{figure}[!htp]
    \centering
    \includegraphics[width=0.5\textwidth]{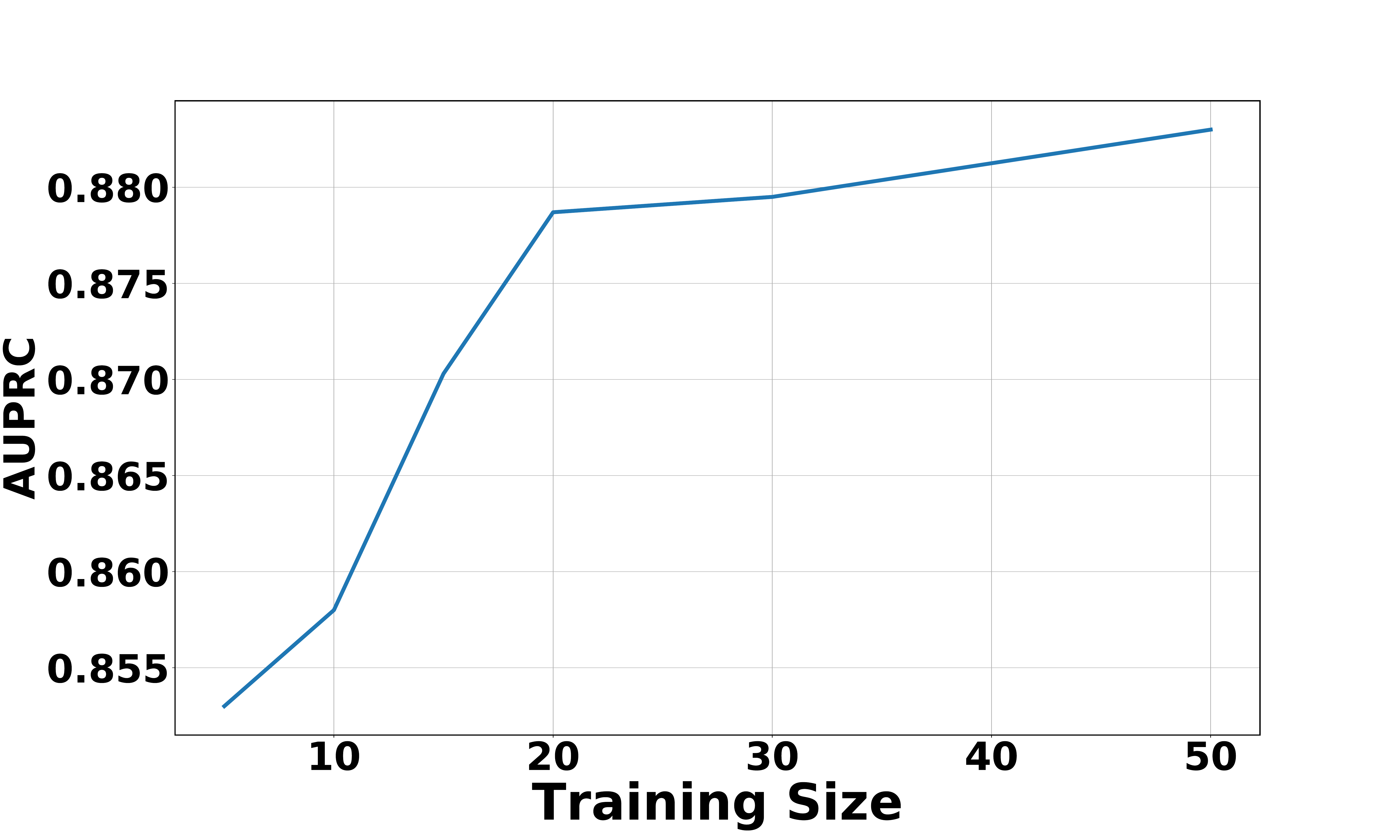}
    \caption{AUPRC using transfer learning in {\em intra-category setup} with gradual increase in retraining percentages. Similar trends are obtained with the other setups.}  
    \label{fig:dicuss}
\end{figure}

 
 \noindent\textbf{Model retraining:}  We demonstrate in our experiments that a fraction of edits from unseen pages results in the improvement over pretrained models. We further investigate the model performance if we increase the volume of the retraining data (results shown for the intra-category setup, all other setups show exactly similar trend). We vary the unseen data used for fine tuning the model from 5\% to 50\% and show that growth in AUPRC stabilizes (see Fig~\ref{fig:dicuss}) which validates our proposal to utilize a smaller fraction. 

 \noindent\textbf{Anecdotal examples}: In order to obtain a deeper understanding of the results, we explore few examples where the general model fails while retraining the dense layers leads to correct classification. In Table~\ref{tab:transfer_example} we present two such examples. 
 Note that our general model (without retraining the dense layers) wrongly  classifies them as damaging edits while retraining leads to correct classification. 
 We believe that retraining the dense layers leads to obtaining superior representation of edits, whereby, page specific words like `streetmaps', `route planning' in {\it Google Maps} or `privacy complaints', `trusted computing' in {\it Internet Privacy} are more pronounced. 
 

\noindent\textbf{Timing benefits}: Another potential benefit is the amount of time saved per epoch as we are only back propagating through the dense layers. 
To quantify the benefit in terms of time, we select a random sample of pages and train one version of our  model end-to-end across all layers and another version only up to the dense layer. For our model, the average time taken per epoch achieves $\sim$ 5x improvement over the traditional approach. 
The performance in the two cases are almost same. In fact, for some cases the traditional end-to-end training leads to inferior results as LSTM layers fail to learn the best weights with so few examples. 

\if{0}
\begin{table}
   \centering
  \begin{tabular}{lc{1.25in}}
    \toprule
    Model & Avg. time  (sec)\\
    \midrule
    Traditional model\\ (bptt to LSTM layers)  & 64  \\
    Specialized model\\ (bptt to dense layer only) & 13 \\
    \bottomrule
  \end{tabular}
  \caption{Average time taken per epoch during training.}
  \label{tab:wikitime}
\end{table}
\fi


\section{Related work}
{\bf Edit quality prediction in Wikipedia} has mostly been pursued in the lines of vandalism detection. ~\citet{kumar2015vews} developed a system which utilized novel patterns embedded in user editing history, to predict potential vandalism. Similar feature based approach has also been applied in both standard~\cite{green2017spam} and sister projects of Wikipedia  such as wikidata~\cite{heindorf2016vandalism,sarabadani2017building}.~\citet{yuan2017wikipedia} propose to use a modified version of LSTM to solve this problem, hence avoiding feature engineering. A complementary direction of investigation has been undertaken by~\cite{daxenberger2013automatically,bronner2012user} who bring forth a feature driven approach, to distinguish spam edit from a quality edit. A feature learning based approach has been proposed by ~\citet{agrawal2016predicting,yardim2018can} which observes all the past edits of a user to predict the quality of the future edits. Temporal traces generated by edit activity has also been shown~\cite{tabibian2017distilling} to be a key indicator toward estimating reliability of edits and page reputation. One of the major problems in these approaches is that they require user level history information which is difficult to obtain because the same user may edit different Wikipedia pages of diverse categories and it will be time consuming to comb through millions of pages for each user. There has also been no work to understand the possibility of predicting edit quality based on edits in pages in a common category. However, there has been no work to leverage advanced machinery developed in language modeling toward predicting edit quality.  

\noindent{\bf Transfer learning:}
Several works~\cite{long2015learning,sharif2014cnn} in computer vision (CV) focus on transfer learning approach as deep learning architectures in CV tend to learn generic to specific tasks from first to last layer. More recently \cite{long2015fully,donahue2014decaf} have shown that fine tuning the last or several of the last layers and keeping the rest of the layers frozen can have similar benefits. 
In natural language processing (NLP) literature, \cite{severyn2015unitn} showed that unsupervised language model based embedding can be tuned using a distant large corpus and then further applied on a specialized task such as sentiment classification. This approach of {\it weak supervision} followed by {\it full supervision} to learn a {\it confident model}  ~\cite{dehghani2017learning,howard2018universal,jan2016sentiment} has been shown to reduce training times in several NLP tasks. In this paper we apply a similar framework for the first time in predicting the edit quality in Wikipedia pages in one category by initializing parameters from a trained model of a different category. This is very effective in cases where the former category page has limited number of data points.

\section{Conclusion}
In this paper we proposed a novel deep learning based model {\it StRE } for quality prediction of edits in Wikipedia. Our model combines word level as well as character level signals in the orthography of Wikipedia edits for extracting a rich representation of an edit. We validate our model on a novel data set comprising millions of edits and show efficacy of our approach compared to approaches that utilize handcrafted features and event based modelling. One of the remarkable findings of this study is that only 20\% of training data is able to boost the performance of the model by a significant margin. 

To the best of our knowledge, this is the first work which attempts to predict edit quality of a page by learning signals from similar category pages as well as cross category pages. We further show applications of recent advances in transfer learning in this problem and obtain significant improvements in accuracy without compromising training times. We believe this work will usher considerable interest in understanding linguistic patterns in Wikipedia edit history and application of deep models in this domain.  
\bibliography{acl2019}
\bibliographystyle{aaai.bst}

\appendix


\label{sec:supplemental}

\end{document}